\documentclass[letterpaper]{article} % DO NOT CHANGE THIS
\usepackage{aaai25}  % DO NOT CHANGE THIS
\usepackage{times}  % DO NOT CHANGE THIS
\usepackage{helvet}  % DO NOT CHANGE THIS
\usepackage{courier}  % DO NOT CHANGE THIS
\usepackage[hyphens]{url}  % DO NOT CHANGE THIS
\usepackage{graphicx} % DO NOT CHANGE THIS
\usepackage{amsmath}

\usepackage{natbib}  % DO NOT CHANGE THIS AND DO NOT ADD ANY OPTIONS TO IT
\usepackage{caption} % DO NOT CHANGE THIS AND DO NOT ADD ANY OPTIONS TO IT
\usepackage{algorithm}
\usepackage{algorithmic}
\usepackage[nolist]{acronym}
\usepackage{booktabs}
\usepackage{multirow}
\usepackage{makecell}
\usepackage{newfloat}
\usepackage{listings}
\DeclareCaptionStyle{ruled}{labelfont=normalfont,labelsep=colon,strut=off} % DO NOT CHANGE THIS
\floatstyle{ruled}
\newfloat{listing}{tb}{lst}{}
\floatname{listing}{Listing}
\title{FaceSpeak: Expressive and High-Quality Speech Synthesis from  Human Portraits of Different Styles}
\author{
   Tian-Hao Zhang\textsuperscript{\rm 1,}\thanks{Work done during internship at Tencent AI Lab.},
    Jiawei Zhang\textsuperscript{\rm 1},
    Jun Wang\textsuperscript{\rm 2},
    Xinyuan Qian\textsuperscript{\rm 1,}\thanks{ Corresponding author.},
    Xu-Cheng Yin\textsuperscript{\rm 1}
}
\affiliations{
    \textsuperscript{\rm 1}School of Computer and Communication Engineering, University of Science and Technology Beijing, Beijing, China\\
    \textsuperscript{\rm 2}Tencent AI Lab, Shenzhen, China\\
    tianhaozhang@xs.ustb.edu.cn, jiaweizhang@xs.ustb.edu.cn
}

\begin{document}
\newacro{TTS}[TTS]{Text-to-speech}
\newacro{EMTTS}[EM$^2$TTS]{Expressive Multi-Modal TTS}
\newacro{PIV}[PIV]{Potrait-insightVoice}
\newacro{sota}[SOTA]{state-of-the-art}
\newacro{LLM}[LLM]{Large Language Model}
\newacro{DSP}[DSP]{Diverse Speaker Prediction}
\newacro{GRL}[GRL]{Gradient Reverse Layer}
\newacro{MI}[MI]{Mutual information}

\maketitle

\begin{abstract}
Humans can perceive speakers' characteristics (e.g., identity, gender, personality and emotion) by their appearance, which are generally aligned to their voice style. Recently, vision-driven \ac{TTS} scholars grounded their investigations on real-person faces, thereby restricting effective speech synthesis from applying to vast potential usage scenarios with diverse characters and image styles. To solve this issue, we introduce a novel FaceSpeak approach. It extracts salient identity characteristics and emotional representations from a wide variety of image styles. Meanwhile, it mitigates the extraneous information (e.g., background, clothing, and hair color, etc.), resulting in synthesized speech closely aligned with a character's persona. Furthermore, to overcome the scarcity of multi-modal \ac{TTS} data,  we have devised an innovative dataset, namely~\ac{EMTTS}, which is diligently curated and annotated to facilitate research in this domain. The experimental results demonstrate our proposed FaceSpeak can generate portrait-aligned voice with satisfactory naturalness and quality. 
% Demos are released at https://facespeak.github.io/.~\footnote{Code and ~\ac{EMTTS} dataset will be public once accepted.}
\end{abstract}
\begin{links}
    \link{Demos}{https://facespeak.github.io}
    % \link{Datasets}{https://aaai.org/example/datasets}
    % \link{Extended version}{https://aaai.org/example/extended-version}
\end{links}

\section{Introduction}

Human voices contain munificent information in aspects such as age~\cite{grzybowska2016speaker,singh2016relationship}, gender~\cite{li2019improving},  emotional nuances~\cite{wang2017learning,zhang2019attention}, physical fitness~\cite{verde2021covid}, and speaker identity~\cite{deaton2010understanding,ravanelli2018speaker}. These vocal characteristics are intrinsically related to an individual's physical and psychological makeup~\cite{xu2024facechain,hardcastle2012handbook}, offering a unique profile of the speaker. For example, the emotional content conveyed through speech is often mirrored in facial expressions. This inherent correlation between voice and image sparked research exploration in various fields,  including emotion recognition~\cite{zhou2021information,lei2023audio,zhang2021continuous}, speaker verification~\cite{qian2021audio,nawaz2021cross}, face-speech retrieval~\cite{li2023rethinking}, and speech separation~\cite{gao2021visualvoice,lee2021looking}. 

\begin{figure}[!tb]
  \centering
  \includegraphics[width=1\columnwidth]{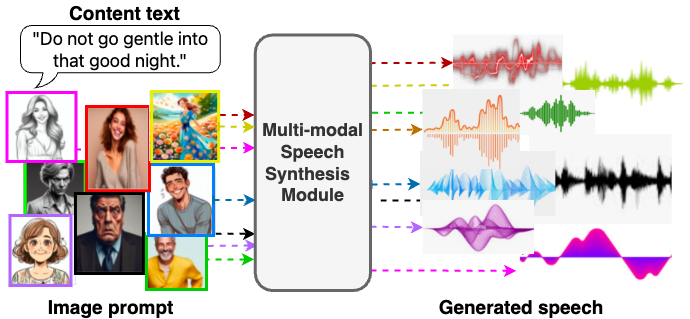}
  \caption{Our proposed multi-modal speech synthesis framework, namely FaceSpeak, which performs expressive and high-quality speech synthesis, given \textit{image prompt} of different styles and the \textit{content text} (Note: image-speech data from various characters are encoded with distinct color codecs).}
  \label{fig:introduction}
 \end{figure}    

% \begin{table*}[!th]
% \centering
% \small
%     \begin{tabular}{p{5.6cm}ccc}
%         \toprule   
%         % \toprule
%         Method & Dataset & Prompt & Backbone \\ \midrule
%         Meta-StyleSpeech~\cite{min2021meta}& VCTK, LibriTTS&Speech&FastSpeech2\\  
%         YourTTS~\cite{casanova2022yourtts}&VCTK, LibriTTS, MLS-PT &Speech&VITS\\ 
%         GenerSpeech~\cite{huang2022generspeech}&IEMOCAP, LibriSpeech, VoxCeleb1, ESD&Speech&Transformer\\  
%          Mega-TTS~\cite{jiang2023mega}&GigaSpeech, WenetSpeech, VCTK, LibriSpeech&Speech&GAN \\  \midrule
%         PromptTTS~\cite{guo2023prompttts}&PromptSpeech&Text&Transformer\\  
%         InstructTTS~\cite{yang2023instructtts}&NLSpeech&Text&Diffusion\\  
%         Sally~\cite{ji2024textrolspeech}&TextrolSpeech&Text&Codec-LM \\ 
%         Promptspeaker~\cite{zhang2023promptspeaker}&Internal Stylistic, AISHELL-3, DiDiSpeech&Text&VITS \\ \midrule VisualTTS~\cite{lu2022visualtts}&GRID&Face&Tacotron \\   
%         Imaginary Voice~\cite{lee2023imaginary} & LRS3, LJSpeech&Face&Diffusion\\ 
%         MMTTS~\cite{guan2024MMTTS}& LJSpeech, MEAD-TTS & Speech,Text,Face & FastSpeech2\\ \midrule
%         \textbf{FaceSpeak} & \textbf{EM$^2$TTS}  & Free-style Image & VITS2 \\
%         \bottomrule 
%                 % \bottomrule
%     \end{tabular}
%     \caption{Summary of the existing TTS systems given \textit{speech}, \textit{face} and \textit{text} as the prompts.} 
%     \label{tab:sota}
% \end{table*}

In recent years, growing research interest has focused on the controllability of synthesized speeches. To this end, one solution is to introduce an auxiliary reference input to model the style features of speech. 
For example, PromptTTS~\cite{guo2023prompttts} and InstructTTS~\cite{yang2023instructtts} 
leverage textual descriptions to control the speech synthesis style.
In contrast, the input text description still needs human crafting efforts and expertise, while some individuals may struggle to accurately express their intended synthesis goals. 
Other works employ vision as a reference. For example, visualTTS~\cite{lu2022visualtts} generates temporal synchronized speech sequences for visual dubbing, and MM-TTS~\cite{guan2024MMTTS} transfers multi-modal prompts (i.e., text, human face, pre-recorded speech) into a unified style representation to control the generation process. Despite both works using images (real human faces) as the reference, thus cannot adapt to nonphotorealistic portraits which are widespread in digital assistants, video games, and virtual reality scenarios. 

Although previous attempts have been made to generate speech based on visual cues, they have faced notable limitations. One such limitation is the predominant reliance on real-face datasets, which lack diversity in image styles necessary for comprehensive speech synthesis guidance. This significantly restricts the potential applications of synthesizing speech from portrait images. Additionally, prior methods often employ entangled embeddings of facial images to guide speech synthesis, potentially introducing extraneous information that hampers performance. Moreover, relying on entangled visual features can further constrain the flexibility of the synthesis system, as the synthesized speech in this case can only be controlled by a single image. However, decoupling identity and emotion features can enable the control of speech synthesis by using different images providing identity and emotion information separately, greatly increasing the diversity and flexibility of multi-modal \ac{TTS}.

In this paper, we tackle the aforementioned issues through a novel multi-modal speech synthesis process. As shown in Fig.~\ref{fig:introduction}, given the text of the content and the images in different styles (e.g., fantasy art and cartoon), our aim is to generate high-quality vivid human speech that is aligned with the characteristics indicated by vision. Our key contributions are summarized as follows. 

\begin{enumerate}
    \item We introduce \ac{EMTTS}, a pioneering multi-style, multi-modal \ac{TTS} dataset. It is designed and re-annotated through a collaborative multi-agent framework which leverages chatGPT\footnote{https://chat.openai.com/} for crafting intermediate textual descriptions, PhotoMaker~\cite{li2023photomaker} for generating human portraits from text, and DALL-E to establish multi-modal coherence.  It provides large-scale and diverse style images that enable the training model to generate high-quality and image-coherent speech, thus facilitates the \ac{sota} multi-modal \ac{TTS} development.
     \item We propose a novel speech synthesis method given human portrait prompt, namely FaceSpeak, to generate speech that is aligned with the characteristics indicated by the visual input. To our best knowledge, this is the first multi-modal expressive speech synthesis work that allows input any-style images. In particular, we disentangle the identity and expression features from facial images, ensuring that the synthesized audio aligns with the speaker's characteristics, while mitigating the impact of irrelevant factors in the images and enhancing the flexibility and diversity of synthesis systems.
    \item Extensive experiments demonstrate that our proposed FaceSpeak can synthesize image-aligned, high-quality, diverse, and expressive human speech. Its superior performance is also validated through the numerous subjective and objective evaluations.
\end{enumerate}

\section{Related Work}
Existing \ac{TTS} works utilizing prompts leverage a multitude of modalities, including reference speech, textual descriptions, and human faces.
% We list the most related works in Table~\ref{tab:sota}.

\textbf{Speech prompt}:
Traditional \ac{TTS} system extracts features from a reference speech to obtain the desired voice with unique vocal characteristics. For example, Meta-StyleSpeech~\cite{min2021meta}, which is built on FastSpeech2~\cite{ren2020fastspeech}, fine-tunes the gain and bias of textual input based on stylistic elements extracted from a speech reference, facilitating effective style-transferred speech synthesis.
YourTTS~\cite{casanova2022yourtts}, based on VITS, proposes modifications for zero-shot multi-speaker and multilingual training, resulting in good speaker similarity and speech quality.
GenerSpeech~\cite{huang2022generspeech} proposes a multi-level style adapter and a generalizable content adapter to efficiently model style information. 
Mega-TTS~\cite{jiang2023mega} employs various techniques (e.g., VQ-GAN, codec-LM) to extract different speech attributes (e.g., content, timbre, prosody, and phase), leading to successful speech disentanglement. 
Despite the impressive results, they are still limited by the availability of pre-recorded reference speech with a clean background. Moreover, the synthesized speech is often limited by the intrinsic attributes of the reference speech. 

\textbf{Text prompt}: 
In contrast to traditional TTS systems that require users to have acoustic knowledge to understand style elements such as prosody and pitch, the use of text prompts is more user-friendly as text descriptions offer a more intuitive and natural means of expressing speech style.
For example, PromptTTS~\cite{guo2023prompttts} utilizes the BERT model as a style encoder and a transformer-based content encoder to extract the corresponding representations from the text prompt to achieve voice control.
InstructTTS~\cite{yang2023instructtts} proposes a novel three-stage training procedure to obtain a robust sentence embedding model that can effectively capture semantic information from style prompts and control the speaking style in the generated speech.
Sally~\cite{ji2024textrolspeech} employs an autoregressive codec-LM as a style encoder and a non-autoregressive codec-LM as a decoder to generate acoustic tokens across varying granularities, capitalizing on the robust coding capabilities of language models.
Promptspeaker~\cite{zhang2023promptspeaker} integrates the Glow model to create an invertible mapping between semantic and speaker representations, thereby enabling text-driven \ac{TTS}.
Despite  promising results, these works still rely on human effort to provide detailed text descriptions, which may be unavailable at scenarios requiring rapid content creation.

\textbf{Image prompt}: 
Image prompt-based TTS allows a more comprehensive and expressive synthesis process, as the visual context can provide additional information and nuances that enhance the overall quality and authenticity of the generated speech.
For example, VisualTTS~\cite{lu2022visualtts} pioneers the use of silent pre-recorded videos as conditioned inputs to not only generate human-like speech but also achieve precise lip-speech synchronization.
Similarly, Imaginary Voice~\cite{lee2023imaginary} introduces a face-styled diffusion \ac{TTS} model within a unified framework which
designs a speaker feature binding loss to enforce similarity between the generated and real speech segments in the speaker embedding space.
To be noted, MMTTS~\cite{guan2024MMTTS} proposes an aligned multi-modal prompt encoder that embeds three different modalities into a unified style space. Despite allowing any modality input, it limits speech generation to real faces, overlooking the potential influence of images with diverse styles. 

\begin{figure*}[!th]
  \centering
  \includegraphics[width=2.0\columnwidth]{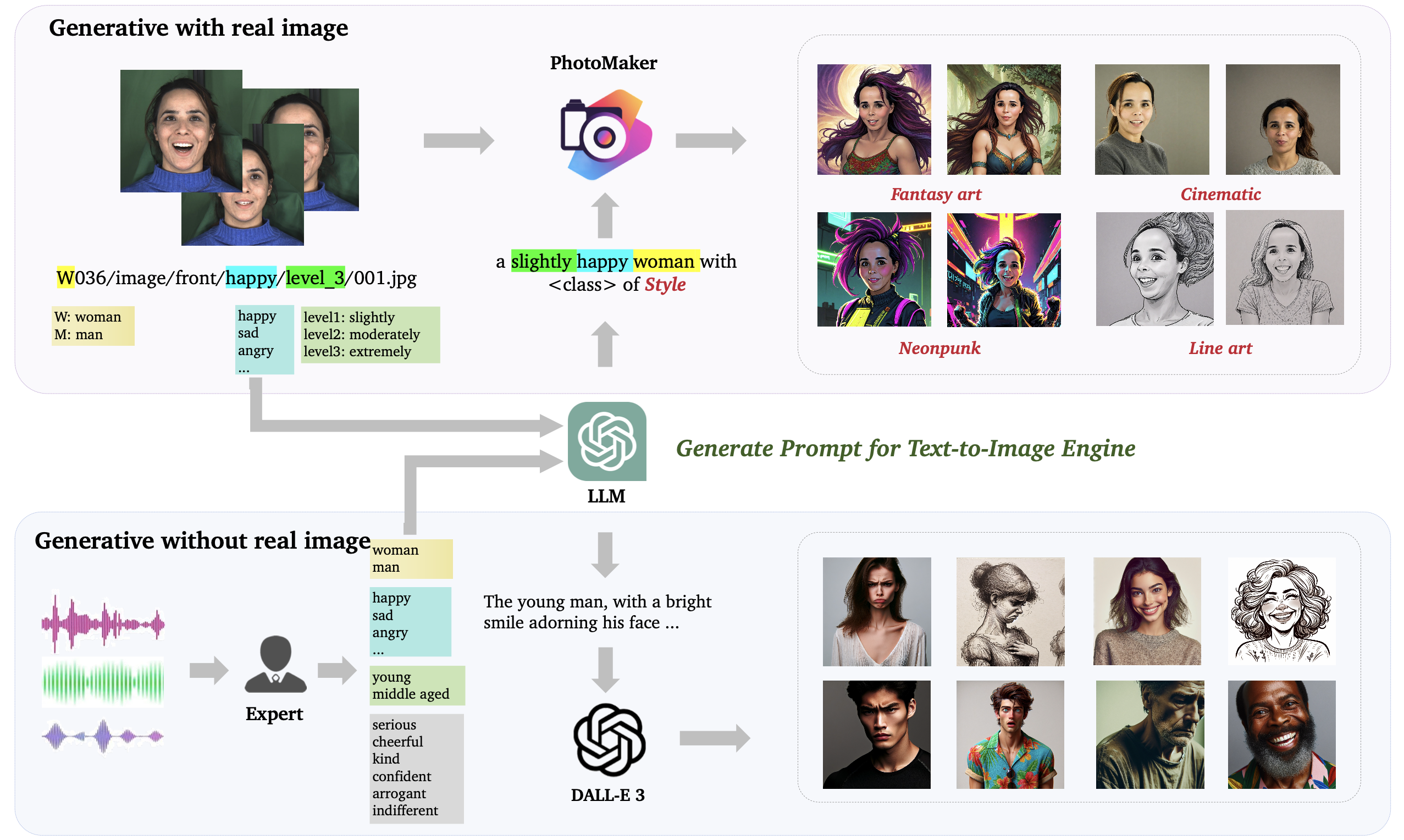}
  \caption{Our image generation pipeline of 
  1) {\ac{EMTTS}-MEAD} subset (top): we specify the desired output style and transfer the real human image to images of different styles using PhotoMaker.  2) {\ac{EMTTS}-ESD-EmovDB} subset (bottom): we use a \textit{human expert} to label the character factors for chatGPT to create the descriptive text, which is  utilized by  DALL-E-3 to  produce images that are highly aligned with the specified parameters. }
  \label{fig:image_pipeline}
\end{figure*}

\textbf{Summary}: Considering the aforementioned limitations, in this paper, we aim to generate expressive and high-quality human speech associated with characters from input images of various styles. Henceforth, we introduce a portrait-speech pairing dataset comprising multi-style images. Using this dataset, we develop FaceSpeak, enabling the model to be generalized across various image styles. In particular, our FaceSpeak framework achieves enhanced accuracy and flexibility in controlling visual features for synthesized speech, which is achieved through the deliberate decoupling of identity and emotion information within the visual features of the portrait.

\section{Proposed \ac{EMTTS} Dataset}\label{sec:EMTTSdataset}
The widely-used TTS datasets, such as LibriTTS~\cite{zen2019libritts} and VCTK~\cite{yamagishi2019cstr}, predominantly exhibit a single-modal nature, comprising solely audio recordings without corresponding textual or visual labels.
Existing multi-modal TTS datasets, including ESD~\cite{zhou2022emotional}, EmovDB\footnote{https://www.openslr.org/115}, and Expresso~\cite{nguyen2023expresso}, lack visual labels for character profiles. While IEMOCAP~\cite{busso2008iemocap}, MEAD~\cite{wang2020mead}, CMU-MOSEI~\cite{zadeh2018multimodal} and RAVDESS~\cite{livingstone2018ryerson} datasets provide labels with limited aspects such as emotion and facial expressions, no existing \ac{TTS} datasets provide face data with diverse  image styles. 

To address the above limitations, we re-design and annotate an expansive multi-modal TTS dataset, termed \ac{EMTTS}. It is enriched with diverse text descriptions and a wide range of facial imagery styles. Due to the distinct data characteristics and varying levels of annotation completion, we have delineated the dataset into two distinct emotional subsets. Please see \textit{Appendix} for more details.

\subsection{\ac{EMTTS}-MEAD}
For the MEAD dataset~\cite{wang2020mead} with real human face recordings, we initially applied a random selection strategy to extract video frames. Then, following steps are designed to proceed the data:
1) \textit{Automatic text generation}:
we generate corresponding text labels that encompass gender, emotion and its intensity, utilizing the raw data from the MEAD dataset.
In particular, our methodology draws from the MMTTS framework~\cite{guan2024MMTTS}, which correlates emotion intensity levels with specific degree words, as shown in Figure ~\ref{fig:image_pipeline}.
2) \textit{Image style transfer}: we leverage the innovative image style conversion model, PhotoMaker~\cite{li2023photomaker}, which excels in producing a variety of portrait styles representing the same individual, guided by character images and textual prompts. For each speaker exhibiting varying emotion intensities, we have created four styles of images (i.e., fantasy art, cinematic, neonpunk, and line art), enriching the visual diversity of the dataset.

Nevertheless, a significant challenge arises in discerning the correlation between a speaker's appearance and their timbre, particularly with \textit{hard samples}.
These cases involve a mismatch between the physical attributes and the vocal characteristics of the speaker (e.g., a physically strong person's voice may sound similar to that of a slim woman).
Consequently, models trained exclusively on \ac{EMTTS}-MEAD face difficulties in aligning with intuitive expectations of human perception. 

\begin{figure*}[tb]
  \centering
  \includegraphics[width=2.1\columnwidth]{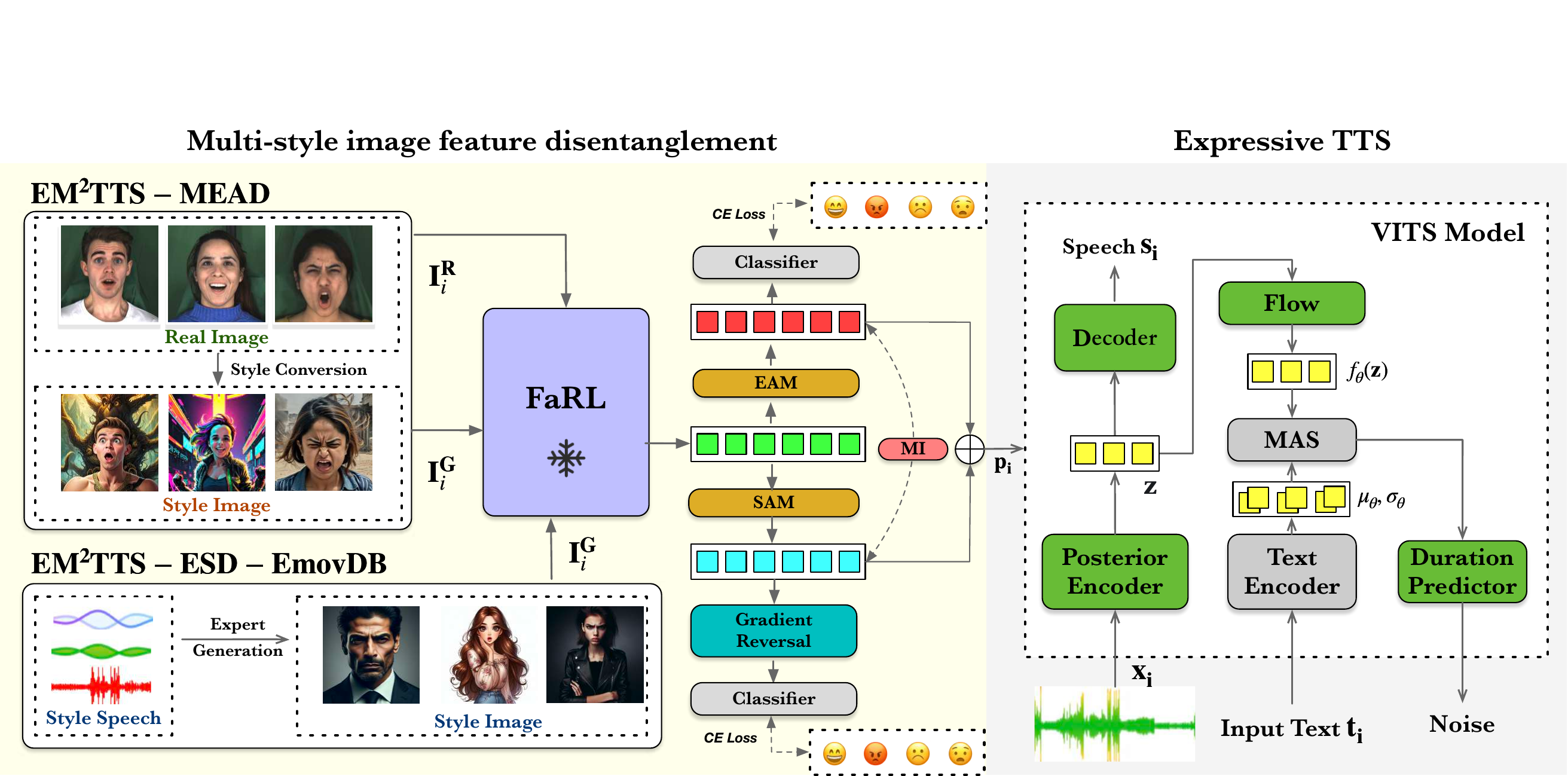}
  \caption{Block diagram of our proposed FaceSpeak which generates speech  given the input text ${\bf t}_i$ and images of different styles (either real ${\bf I}^R_i$ or generated ${\bf I}^G_i$).
  It consists of two sub-modules: multi-style image feature disentanglement (yellow region) and expressive TTS (gray region).}
  \label{fig:method}
\end{figure*}

\subsection{\ac{EMTTS}-ESD-EmovDB}
The unimodal emotional speech datasets ESD and EmovDB lack accompanying speaker images, thus performing style transfer based on real human face is unfeasible.
Therefore, we design the next steps to process them:
1) \textit{Manual annotation}: we explore a human expert to label age, gender, and characteristics by listening to each speech data; 2) \textit{Text expansion}: we use the \ac{LLM} model e.g., ChatGPT to expand the label words into texts with varying contents but similar meanings; 3) \textit{Text-driven image generation}: the enriched texts were then fed into DALL-E-3, a text-to-image model capable of generating a multitude of images in distinct styles. 
We assigned these generated images to the corresponding speeches, resulting in style-free images that emphasize the character's emotions, surpassing the limitations of the recorded images.

\section{Proposed Method}
Let us denote ${\bf I}_{i}$, ${\bf x}_{i}$ and ${\bf t}_i$ as the visual image of arbitrary style, the corresponding voice signal, and the co-speech content of character $i$, respectively.
As depicted in Fig.~\ref{fig:method}, our proposed FaceSpeak algorithm aims to generate a speech waveform ${\bf s}_{i}$ corresponding to the imagined voices of characters given the input text ${\bf t}_i$ and images of different styles (either real ${\bf I}^R_{i}$ or generated ${\bf I}^G_{i}$).

The proposed FaceSpeak consists of two sub-modules: 1) Multi-style image feature decomposition module that receives different styles of portrait images for feature extraction and decouples emotion and speaker information. With this module, we can get the disentangled identity and emotion embeddings extracted from the portrait visual features; 2) Expressive TTS module receives the identity and emotion embeddings as control vectors for generating high-quality speech that matches the portrait images. Detailed descriptions are given below.

\subsection{Multi-Style Image Feature Disentanglement}
To enhance the coherence of identity and emotion between the input image and the synthesized speech, it is crucial to mitigate the influence of extraneous visual elements (e.g., background, clothing, distracting objects) in the extracted visual features.
The FaRL model~\cite{zheng2021general}, leveraging the multi-modal pre-trained CLIP model on large-scale datasets of face images and correlated text, 
ensures the extraction of predominantly face-related visual features with robust generalization capabilities.
Thus, we apply it on real images and corresponding multi-style images to extract high-dimensional intermediate visual representations:
\begin{equation}
    \textbf{e}_i=\mathrm{FaRL}(\bf{I}_i)
\end{equation}
where $\textbf{e}_i \in \mathbf{R}^{512}$ contains both the emotion and speaker information of the portrait. 

Our subsequent objective is to decouple the emotion and the speaker information within $\textbf{e}_i$. 
Let us define the Identity Adapter Module (IAM) and the Expression Adapter Module (EAM) to learn the mapping from  $\textbf{e}_i$ to the identity embedding  $\boldsymbol{\alpha}_i$ and emotion embedding $\boldsymbol{\beta}_i$, respectively:
\begin{equation}
    \begin{split}
        \boldsymbol{\alpha}_i=\mathrm{IAM}(\textbf{e}_i)= \mathrm{FC} \left (\mathrm{GeLU}(\mathrm{FC}(\textbf{e}_i)) \right )\\
        \boldsymbol{\beta}_i=\mathrm{EAM}(\textbf{e}_i) = \mathrm{FC}\left (\mathrm{GeLU}(\mathrm{FC}(\textbf{e}_i)) \right )
    \end{split}
\end{equation}

Subsequently, we used the emotion classification model following $\boldsymbol{\beta}_i$ to bias the $\boldsymbol{\beta}_i$ features toward emotion characteristics. Furthermore, we incorporated the emotion classification model after introducing \ac{GRL} following $\boldsymbol{\alpha}_i$, which aims to minimize the sentiment information retained in $\boldsymbol{\alpha}_i$. We utilize the Cross-Entropy loss to constrain the two emotion classification models, formulated as follows:
\begin{equation}
\small
    \begin{split}
        \mathcal{L}_{emo} &= \mathrm{CrossEntropy}(\mathrm{CLS}(\boldsymbol{\beta}_i), \mathrm{L}_e) \\
        \mathcal{L}_{grl} &= \mathrm{CrossEntropy}(\mathrm{GRL}(\mathrm{CLS}(\boldsymbol{\alpha}_i)), \mathrm{L}_e) \\
    \end{split}
\end{equation}
where $\mathrm{CLS}$ represents the classification layer,
% whose output dimension is the number of emotional categories,
and $\mathrm{L}_e$ denotes the emotion categorization label of the input image $\bf{I}_i$. $\mathrm{GRL}$  inverts the sign of the incoming gradient during the back-propagation phase. By this strategic reversal, $\mathrm{IAM}$ learns to remove or minimize features that are correlated with emotion, emphasizing the identity aspects of the input.

To enhance the decoupling of identity embedding $\boldsymbol{\alpha}_i$ and emotion embedding $\boldsymbol{\beta}_i$, we propose a feature decoupling approach that leverages \ac{MI} minimization. This strategy effectively reduces the correlation between identity and emotion representations, enabling a more robust and accurate analysis of each aspect.

$\textbf{Mutual information based decoupling}$: \ac{MI} is a fundamental concept in information theory that quantifies the statistical dependence between two random variables $V_1$ and $V_2$, calculated as:
\begin{equation}
\small
    \begin{split}
    I(V_1; V_2) = \sum_{v_1 \in V_1} \sum_{v_2 \in V_2} p(v_1, v_2) \log\left(\frac{p(v_1, v_2)}{p(v_1)p(v_2)}\right)
\end{split}
\end{equation}
where $p(v_1, v_2)$ is the joint probability distribution between
$v_1$ and $v_2$, while $p(v_1)$ and $p(v_2)$ are their marginals. However, it is still a challenge to obtain a differentiable and scalable MI estimation. In this work, we use vCLUB \cite{DBLP:conf/icml/ChengHDLGC20}, an extension of CLUB, to estimate an upper bound on \ac{MI}, which allows efficient estimation and optimization of MI when only sample data are available and a probability distribution is not directly available. Given the sample pairs of identity embedding and emotion embedding $\{(\boldsymbol{\alpha}_i, \boldsymbol{\beta}_i)\}_{i=1}^N$, where $N$ denotes the number of samples, the MI can be computed as:
\begin{equation}
\small
    \label{mi_loss}
    \begin{split}
\mathcal{L}_{mi} = \frac{1}{N^2} \sum_{i=1}^{N} \sum_{j=1}^{N} \left[ \log q_\theta(\boldsymbol{\beta}_i|\boldsymbol{\alpha}_i) - \log q_\theta(\boldsymbol{\beta}_j|\boldsymbol{\alpha}_i) \right]
\end{split}
\end{equation}
where $q_\theta(\boldsymbol{\beta}_i|\boldsymbol{\alpha}_i)$ is a variational approximation which can make vCLUB holds a MI upper bound or become a reliable MI estimator. At each iteration during the training stage, we first obtain a batch of samples $\{(\boldsymbol{\alpha}_i, \boldsymbol{\beta}_i)\}$  from $\mathrm{IAM}$ and $\mathrm{EAM}$, then update the variational approximation $q_\theta(\boldsymbol{\beta}_i|\boldsymbol{\alpha}_i)$ by maximizing the log-likelihood $\mathcal{L}_{\theta} = \frac{1}{N} \sum_{i=1}^{N} \log q_\theta(\boldsymbol{\beta}_i|\boldsymbol{\alpha}_i)$.
% \begin{equation}
%     \begin{split}
% \mathcal{L}_{\theta} = \frac{1}{N} \sum_{i=1}^{N} \log q_\theta(\boldsymbol{\beta}_i|\boldsymbol{\alpha}_i)
% \end{split}
% \end{equation}
 The updated $q_\theta(\boldsymbol{\beta}_i|\boldsymbol{\alpha}_i)$ can be used to calculate the vCLUB estimator. Finally, we sum the decoupled identity embedding and the emotion embedding obtained as the ultimate control embedding $p_i$ for speech synthesis.

\begin{table*}[th]
    \centering
    \small
    \setlength{\tabcolsep}{5.8mm}{
    \begin{tabular}{lcccccc}
    \toprule
    % \toprule
      Method & \thead{Intra-domain \\ NMOS $\uparrow$ \hspace {0.6cm} ISMOS $\uparrow$  \hspace {0.6cm} ESMOS $\uparrow$ }& \thead{Out-of-domain \\ NMOS $\uparrow$  \hspace {0.6cm}ISMOS $\uparrow$ \hspace {0.6cm} ESMOS $\uparrow$}\\
    \midrule
    GT  & 4.42 $\pm$ 0.02  \hspace {0.9cm}  -  \hspace {0.9cm} 4.52 $\pm$ 0.03 &-  \hspace {1.6cm}  -  \hspace {1.6cm} -\\  
    % GT(Mel+vocoder)   & 4.87 $\pm$ 0.17 & 4.87 $\pm$ 0.17\\
    VITS2  &3.55 $\pm$ 0.06  \hspace {0.2cm}  3.68 $\pm$ 0.07  \hspace {0.2cm} 3.38 $\pm$ 0.13& 3.42 $\pm$ 0.05  \hspace {0.2cm}  3.56 $\pm$ 0.10  \hspace {0.2cm} 3.31 $\pm$ 0.09\\
    MM-StyleSpeech   & 3.58 $\pm$ 0.08  \hspace {0.2cm}  3.64 $\pm$ 0.04  \hspace {0.2cm} 3.89 $\pm$ 0.11 & 3.23 $\pm$ 0.08  \hspace {0.2cm}  3.61 $\pm$ 0.07  \hspace {0.2cm} 3.78 $\pm$ 0.08\\
    MM-TTS   & 3.94 $\pm$ 0.05 \hspace {0.2cm} 3.82 $\pm$ 0.08 \hspace {0.2cm} 4.08 $\pm$ 0.08 & 3.41 $\pm$ 0.06  \hspace {0.2cm}  3.68 $\pm$ 0.04  \hspace {0.2cm} 3.91 $\pm$ 0.05\\
    \textbf{FaceSpeak}  & 4.13 $\pm$ 0.04 \hspace {0.2cm} 3.97 $\pm$ 0.07 \hspace {0.2cm} 4.36 $\pm$ 0.05 & 4.28 $\pm$ 0.05 \hspace {0.2cm} 3.77 $\pm$ 0.09 \hspace {0.2cm} 3.98 $\pm$ 0.07\\
    \bottomrule
    \end{tabular}}
    \caption{MOS results with 95$\%$ confidence interval (N-: naturalness; IS-: identity similarity; ES-: emotion similarity;  -: information not applicable.)}
    \label{tab:MOS}
\end{table*}

\begin{table*}[htbp]
    \centering
    \small
    \setlength{\tabcolsep}{3.4mm}{
    \begin{tabular}{lcccc}
    \toprule
      \multirow{2}*{Method} & \multicolumn{2}{c}{Intra-domain}& \multicolumn{2}{c}{Out-of-domain}\\
      ~ & 7-point score $\uparrow$ &  \thead{Preference ($\%$) \\B \hspace {0.8cm} E \hspace {0.8cm} O} & 7-point score $\uparrow$ & \thead{Preference ($\%$) \\B \hspace {0.8cm} E \hspace {0.8cm} O}\\
    \midrule
    MM-StyleSpeech  & 1.25 $\pm$ 0.08 & 28   \hspace {0.7cm} 22  \hspace {0.7cm} 50 &  1.97 $\pm$ 0.12 & 10   \hspace {0.6cm} 15  \hspace {0.8cm} 75\\
    MM-TTS   & 1.03 $\pm$ 0.06 & 33   \hspace {0.7cm} 14   \hspace {0.7cm} 53 & 1.42 $\pm$ 0.09 & 16   \hspace {0.6cm} 24   \hspace {0.8cm} 60\\
    \bottomrule
    \end{tabular}}
    \caption{AXY preference test results.
B, E and O respect the preference rate for baseline model, equivalent and our model, respectively.}
    \label{tab:AXY}
\end{table*}

\begin{table*}[tb]
\centering
\small
\begin{tabular}{ccccccc}
\toprule
\multirow{2}*{Method} & \multicolumn{4}{c}{Intro-domain} & \multicolumn{2}{c}{Out-of-domain} \\
 & MCD $\downarrow$  & $\mathrm{ACC_{emo}}$ $\uparrow$ & $\mathrm{ACC_{gen}}$ $\uparrow$ & SS $\uparrow$ & $\mathrm{ACC_{emo}}$ $\uparrow$ & $\mathrm{ACC_{gen}}$ $\uparrow$ \\
\midrule
GT & - & 84.54 & 100.00 & - & - & - \\
\textbf{FaceSpeak} & 3.32 & 60.92 & 99.40 & 0.95 & 31.32 & 92.42 \\
\bottomrule
\end{tabular}
\caption{Subjective results on real portraits controlled speech synthesis.}
\label{tab:obj}
\end{table*}

\begin{table*}[hbtp]
\centering
\small
\setlength{\tabcolsep}{6.4mm}{
\begin{tabular}{lcccccc}
\toprule
  Method & NMOS $\uparrow$ & ISMOS $\uparrow$ & ESMOS $\uparrow$ & $\mathrm{ACC_{emo}}$ $\uparrow$ & $\mathrm{ACC_{gen}}$ $\uparrow$  \\
\midrule
w/o EM$^2$TTS & 4.31 $\pm$ 0.05 & 3.88 $\pm$ 0.09 & 4.02 $\pm$ 0.06 & 18.34 & 84.24  \\
w/ EM$^2$TTS  & 4.38 $\pm$ 0.06 & 4.06 $\pm$ 0.08 & 4.47 $\pm$ 0.04 & 26.56 & 92.22  \\
\bottomrule
\end{tabular}}
\caption{Objective and Subjective results on out-of-domain multi-style portraits controlled.}
\label{tab:data-ab}
\end{table*}

% \subsection{Problem formulation}
\subsection{Expressive \ac{TTS}}
We use VITS2~\cite{kong2023vits2}, one of the \ac{sota} \ac{TTS} models, as our speech synthesis backbone. As shown in Figure \ref{fig:method}, The VITS2 model consists of a Posterior Encoder and a Text Encoder that generate posterior distribution and prior distribution based on the input speech and text, respectively; a transformer-based Flow module to refine the latent representation produced by the posterior encoder; a Monotonic Alignment Search (MAS) module to estimate an alignment between input text and target speech; a Duration Predictor module to predict the duration of each phoneme; a Decoder to reconstructs the speech from the latent representation generated by the posterior encoder. In particular, we injected the control embedding from the portrait images into the Posterior Encoder, Decoder, Flow module, and Duration Predictor to generate the speech corresponding to the portrait images, which are highlighted in green in Figure \ref{fig:method}. In the training stage, identity embedding $\boldsymbol{\alpha}_i$ and emotion embedding $\boldsymbol{\beta}_i$ are decoupled from the same image and the final loss can be expressed as:
\begin{equation}
    \begin{split}
        \mathcal{L} = \mathcal{L}_{vits} + \lambda_1\mathcal{L}_{mi} + \lambda_2\mathcal{L}_{emo} + \lambda_3\mathcal{L}_{grl}
    \end{split}
\end{equation}
where $\lambda_1$, $\lambda_2$ and $\lambda_3$ are the hyper-parameters to balance the individual losses. During inference, $\boldsymbol{\alpha}_i$ and $\boldsymbol{\beta}_i$ can come from the same image or be provided by different images separately, and the inference process of Expressive TTS is consistent with work \cite{kong2023vits2}.

\section{Experiments}
\subsection{Dataset and Experimental Setup}
 In the evaluation of our method, we conduct separate experiments on intra-domain and out-of-domain data, respectively. We use \ac{EMTTS}-MEAD as the intra-domain experimental data, while for the out-of-domain evaluation, we perform different settings for real portrait scenes and multi-style virtual portrait scenes, respectively. For the real portrait scenes, we follow the MMTTS's setup which uses the face images in Oulu-CASIA dataset and transcriptions in LibriTTS, while for the multi-style virtual portrait scenario, we use a different image generation API from the training set to generate new test data based on \ac{EMTTS}-ESD. Specifically, we compare our method with the following system:1) GT: Ground Truth. 2) VITS2: A multispeaker TTS baseline. 3) MMTTS: A style transfer TTS system using prompt including image. 4) MM-StyleSpeech: Same as MMTTS using StyleSpeech as backbone. The proposed FaceSpeak is trained for 150K iterations using the Adam optimizer on NVIDIA GeForce RTX 3090 GPUs. Detailed training parameters and network configuration can be found in the \textit{Appendix}. 

\subsection{Synthetic Quality on Real Portraits}
We first evaluate the quality of the speech synthesized by FaceSpeak given a real portrait as the prompt. We conduct a Mean Opinion Score (MOS) ~\cite{Sisman2021MOS} with 95 $\%$ confidence intervals to assess speech quality. Naturalness-MOS (NMOS) and  Similarity-MOS (SMOS) evaluate the speech naturalness and image-speech similarity.
Specifically, we generate 50 speech samples for each model, which are rated by 20 volunteers on a scale of 1 to 5, with higher scores indicating better results ($\uparrow$). We also perform an AXY preference test~\cite{Ryan2018Towards} to verify the style transfer effect based on image prompt. In the test, ``A" is the reference speech that is stylistically consistent with the image prompt, ``X"  and ``Y" are the speech generated by the compared model or our proposed FaceSpeak. The participants decide whether the speech style of ``X" or ``Y" is closer to that of image prompt where the scale range of [-3,3] indicating ``X" is closer to ``Y" is closer. Table \ref{tab:MOS} displays the subjective results including NMOS, ISMOS and ESMOS. Our FaceSpeak achieves better results on both speech naturalness and style similarity, showing the effectiveness of DSP module on speaker representation extraction by using IAM and EAM. As shown in Table \ref{tab:AXY}, the results of AXY test indicate that listeners prefer FaceSpeak synthesis against the compared models. The generated data and the method significantly improves the style extraction ability, allowing an arbitrary reference sample to guide the stylistic synthesis of arbitrary content text.

As shown in the Table \ref{tab:obj}, we further objectively measure the quality of speech through MCD~\cite{Kubichek1993MCD}, emotion and gender classification accuracy, and speaker similarity. MCD measures the spectral distance between the reference and synthesized speech and FaceSpeak achieved a MCD result of 3.32. We measure speaker similarity (SS) between two speech samples in Resemblyzer~\footnote{https://github.com/resemble-ai/Resemblyzer}, our result is 0.95. A hubert-based pre-trained model~\footnote{https://github.com/m3hrdadfi/soxan.git} is used for gender classification ($\text{Acc}_{gen}$) and emotion2vec is used to predict the emotion category ($\text{Acc}_{emo}$) of speech. On the intra-domain data, we obtained $\text{Acc}_{gen}$ for 99.40\% and $\text{Acc}_{emo}$ for 60.92\%. For out-of-domain data, the results for $\text{Acc}_{gen}$ and $\text{Acc}_{emo}$ are 92.42\% and 31.32\%, respectively.

\subsection{Synthetic Quality on Multi-Style Virtual Portraits}
In evaluating the quality of FaceSpeak's speech synthesis based on multi-style virtual portraits, the models we compare are whether or not trained with the virtual portrait images of our proposed multi-style dataset \ac{EMTTS}, respectively. Since models trained with \ac{EMTTS} will have ``seen" multi-style portraits in the domain, our evaluation will only be performed on out-of-domain multi-style portraits. As illustrated in Table \ref{tab:data-ab}, the model trained with the virtual portrait images performs better in all evaluation metrics, confirming the effectiveness of our methods.

\begin{figure}[!tb]  
    \centering
  \includegraphics[width=1.0\columnwidth]{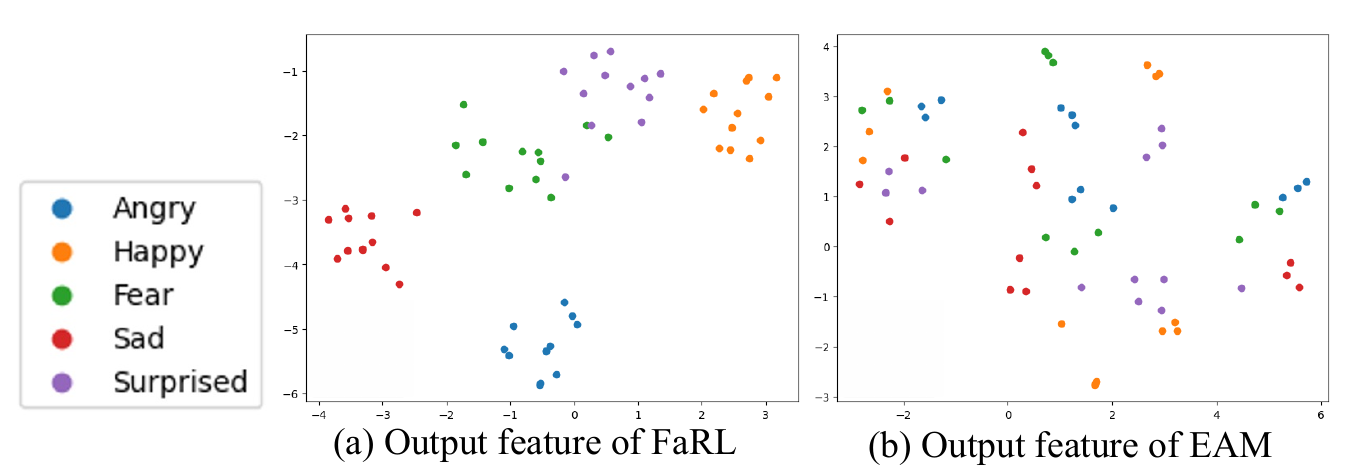}
  \caption{Visualization of emotion embeddings (colors index emotions).}
  \label{fig:emotion_tsne}
\end{figure}

\begin{figure*}[!tb]  
    \centering
  \includegraphics[width=2.1\columnwidth]{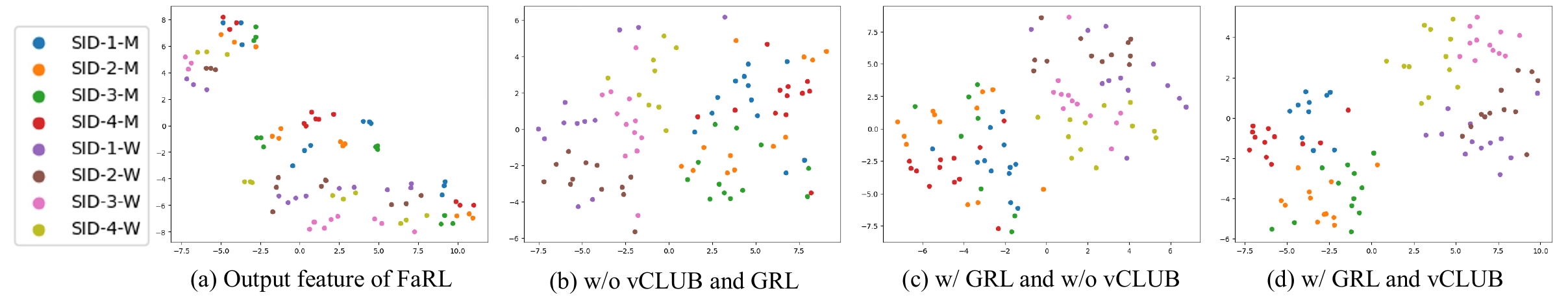}
  \caption{Visualization of identity embeddings (colors index identities; M: male; F: female).}
  \label{fig:sid_tsne}
\end{figure*}

\subsection{Results of Decoupled Identity and Emotion Information}
\textbf{TSNE results of decoupled features}: 
Fig.~\ref{fig:emotion_tsne} (a) visualizes the embeddings extracted from FaRL in emotion, which are randomly distributed. By applying the EAM module, 
as shown in Fig.~\ref{fig:emotion_tsne} (b), the learned embeddings with the same emotion are more clustered, while the others are more discriminative.
Fig.~\ref{fig:sid_tsne} (a) shows the embeddings extracted from FaRL in identity, which are also randomly distributed. By applying the IAM module, from Fig.~\ref{fig:sid_tsne} (b) to (d), it is observed that embeddings from different speakers are more distinguished when using both the GRL and vCLUB strategies.
These intuitively verify the effectiveness of our decoupled identity and emotion characteristics.

\begin{figure}[h]  
    \centering
  \includegraphics[width=0.95\columnwidth]{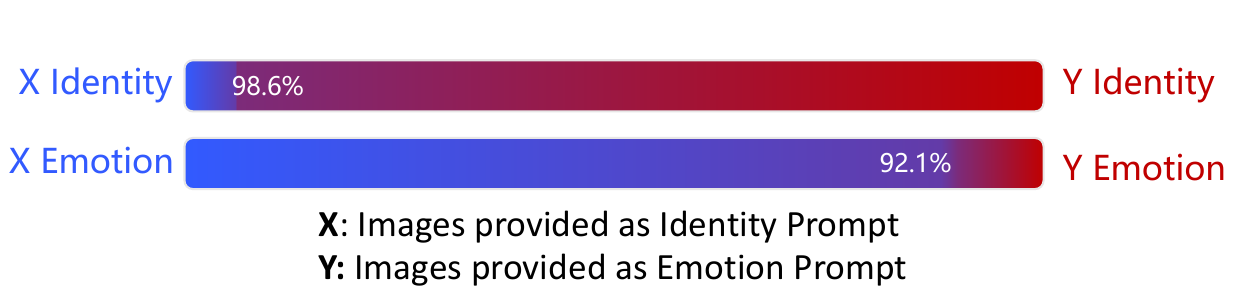}
  \caption{The accuracy in matching the synthesized speech to the emotion and identity styles when they are controlled separately by different images.}
  \label{fig:combined_img}
\end{figure}

\textbf{Speech synthesis controlled by combined portraits}: By decoupling identity and emotion information in an image, we can use different image combinations to control the synthesized speech. As shown in Figure \ref{fig:combined_img}, we define $\mathbf{X}$ as the image that provides identity embedding and  $\mathbf{Y}$ as the image that provides emotion embedding, ensuring that $\mathbf{X}$ and  $\mathbf{Y}$ have different genders and emotions. We let listeners discriminate the synthesized speech by deciding whether it is matched to the  $\mathbf{X}$-image or $\mathbf{Y}$-image in terms of identity and emotion, respectively. As a result, 98.6\% of the speech are determined to correctly match the $\mathbf{X}$ image on identity, while 92.1\% of the speech are determined to correctly match the $\mathbf{Y}$ image on emotion, which proves that our proposed FaceSpeak speech synthesis system can reliably control the synthesis by combining different images, greatly improving the diversity and flexibility.

\section{Conclusion}
In this paper, we introduce FaceSpeak, a pioneering approach for multi-modal speech synthesis which extracts key identity and emotional cues from diverse character images to drive the TTS module to synthesize the corresponding speech. To tackle the problem of data scarcity, we introduced an innovative \ac{EMTTS} dataset, which is meticulously curated and annotated to support and advance research in this emerging field. Additionally, novely methods are proposed for decoupling emotional and speaker-specific features, enhancing both the adaptability and fidelity of our system. Experimental results confirm that FaceSpeak can generate high-quality, natural-sounding speech that authentically align with the visual attributes of the character.

Looking ahead, we aspire to broaden the diversity of speaker categories within the Facespeak system, integrating a wider array of emotions, roles, and other dynamic attributes. By leveraging larger, more comprehensive datasets, we aim to advance the system's development, enhancing its adaptability and versatility.

\section{Acknowledgments}
This work is supported by CCF-Tencent Rhino-Bird Open Research Fund, National Natural Science Foundation of China under Grant No. 62306029, Beijing Natural Science Foundation under Grants L233032, Shenzhen Research Institute of Big Data under Grant No. K00120240007.

\bibliography{aaai25}
\end{document}